\documentclass[%
linenumbers,
10,
jpb,
]{iopart}
\usepackage{graphicx}
\usepackage{hyperref}
\graphicspath{ {./figure/} }
\usepackage{here}
\usepackage{indentfirst}
\usepackage{dcolumn}
\usepackage{bm}
\usepackage{braket}
\usepackage{hyperref}
\usepackage{txfonts}
\usepackage{ulem}
\usepackage{color}

\newcommand{\revise}[1]{\textcolor{red}{#1}}
\newcommand{\ueda}[1]{\textcolor{red}{#1}}
\newcommand{\ota}[1]{\textcolor{green}{#1}}
\newcommand{\yamazaki}[1]{\textcolor{cyan}{#1}}
\newcommand{\hatada}[1]{\textcolor{magenta}{#1}}
\newcommand{\didier}[1]{\textcolor{yellow}{#1}}

\newcommand{\delhatada}[1]{\textcolor{magenta}{\sout{\textcolor{black}{#1}}}}

\renewcommand{\revise}[1]{\textcolor{black}{#1}}
\renewcommand{\ueda}[1]{\textcolor{black}{#1}}
\renewcommand{\ota}[1]{\textcolor{black}{#1}}
\renewcommand{\yamazaki}[1]{\textcolor{black}{#1}}
\renewcommand{\hatada}[1]{\textcolor{black}{#1}}
\renewcommand{\didier}[1]{\textcolor{black}{#1}}

\renewcommand{\delhatada}[1]{\textcolor{black}{\sout{\textcolor{black}{#1}}}}

\begin{document}
\title[\hatada{New formula for $p$- and $s$-wave interference analogous to Young's double-slit experiment for hetero-diatomic molecules}]{Theory of polarization-averaged core-level molecular-frame photoelectron angular distributions: \hatada{III.
\ota{
New formula for $p$- and $s$-wave interference analogous to Young's double-slit experiment for core-level photoemission from hetero-diatomic molecules
}
}
}
\author{
F Ota$^1$, 
K Yamazaki$^{2,3}$
D S\'ebilleau$^4$, 
K Ueda$^5$ and 
K Hatada$^6$
}

\address{$^1$ Graduate School of Science and Engineering for Education, University of Toyama, Gofuku 3190, Toyama 930-8555, Japan}
\address{$^2$ Institute for Materials Research, Tohoku University, 2-1-1 Katahira, Aoba-ku, Sendai 980-8577, Japan}
\address{$^3$ Attosecond Science Research Team, Extreme Photonics Research Group, RIKEN Center for Advanced Photonics, RIKEN, 2-1 Hirosawa, Wako, Saitama, 351-0198, Japan.}
\address{$^4$ D\'epartement Mat\'eriaux Nanosciences, Institut de Physique de Rennes, UMR UR1-CNRS 6251, Universit\'e de Rennes, F-35000 Rennes, France}
\address{$^5$ Department of Chemistry, Graduate School of Science, Tohoku University, 6-3 Aramaki Aza-Aoba, Aoba-ku, Sendai 980-8578, Japan}
\address{$^6$ Faculty of Science, Academic Assembly, University of Toyama, Gofuku 3190, Toyama 930-8555, Japan}

\eads{
\mailto{hatada@sci.u-toyama.ac.jp},
\mailto{oofukiko@gmail.com}}
\date{\today}

\begin{abstract}
We present a new variation of Young's \ueda{double-slit formula} for \ueda{ polarization-averaged molecular-frame photoelectron angular distributions (PA-MFPADs) of hetero-diatomic molecule\didier{s}, which may be used \didier{to extract} the bond length.} 
So far, \ueda{empirical analysis of the \ota{PA-}MFPADs} 
has \ueda{often} been 
carried out \ueda{employing} Young's formula 
\ueda{in which each of \didier{the} two atomic centers emits 
an $s$-photoelectron wave.}
\ueda{The PA-MFPADs, on the other hand, can 
consist of \didier{an} interference between the $p$-wave from the X-ray absorbing atom emitted along the molecular axis and the $s$-wave scattered by neighboring atom, 
\didier{with}in the framework of Multiple Scattering theory.}
\ueda{The difference of this $p$-$s$ wave interference from the commonly used $s$-$s$ wave interference 
causes a \ueda{dramatic} change in the interference pattern, especially near the angles perpendicular to the molecular axis.} \didier{This change involves an additional fringe, urging  us to caution when using the conventional Young's formula for retrieving the bond length.} 
We have derived a new formula \ueda{analogous to Young's formula but for the $p$-$s$ wave interference.}
The bond lengths retrieved from the PA-MFPADs via the new 
formula reproduce the original C-O bond lengths used in the reference \textit{ab-initio} PA-MFPADs within the relative error of 5 \%.
\ota{
In 
the high energy regime, 
\didier{this new} formula for $p$-$s$ wave interference 
\ueda{converges} to the ordinary Young’s formula for the $s$-$s$ wave interference.}
We expect 
\ueda{\didier{it to} be used to retrieve the} bond length \didier{for time-resolved PA-MFPADs} instead of \didier{the} conventional Young's formula.
%
\end{abstract}


\submitto{\jpb}

\maketitle
\ioptwocol

\section{Introduction}

\hatada{
Tracking the real molecular dynamics is one of the main interests in chemistry, biology and physics. It provides us \ota{with} deep insights into \didier{the} properties of molecules, e.g., chemical reactions~\cite{Zewail2000a,Zewail2000b,Zewail2014}.}
%
\hatada{The recent development of X-ray free electron lasers (XFELs)~\cite{Ackermann2007, Emma2010, Ishikawa2012, Allaria2012, Kang2017, Decking2020, Milne2017} has enabled us to investigate the structural changes of molecules and solids in ultra-short-time resolution of the order of femtoseconds, due to \didier{their} high brightness and short pulse width~\cite{Bostedt2016, Seddon2017, Callegari2021}.} 

\yamazaki{Core-level excitation X-ray spectroscopies such as X-ray absorption (XAS) and photoelectron diffraction (PED) are effective approaches for investigating the structure of non-periodic gas\didier{es}, liquid\didier{s} and amorphous systems since they do not require any periodicity.} 
\ueda{Extended X-ray absorption fine structure (EXAFS), a variant of XAS, and PED are conceptually similar: EXAFS scans the amplitude of the photoelectron momentum while PED \hatada{is mostly used for} scans \hatada{of} the direction of the photoelectron momentum and both provide information on the local structure around the core orbital where the X-ray absorption takes place~\cite{Rehr2000}.}
\ueda{PED is commonly used for surface structure analysis~\cite{Woodruff2008}}.

\hatada{In order to apply PED to gas-phase molecules, \yamazaki{the PED scan should be performed in \didier{the} molecular frame.} This is experimentally achieved by angle-resolved coincidence measurements between \didier{the} core-level photoelectrons and \didier{the} fragment ions~\cite{Shigemasa1995}. \didier{The} \yamazaki{COLTRIMS-Reaction microscope~\cite{Ullrich2003} is a standard technique for studying molecular photoionization in the coincidence manner~\cite{Landers2001}:} The COLTRIMS technique  provides the angular distribution of photoelectrons in the molecular frame (Molecular-Frame Photoelectron Angular Distributions: MFPADs), \ueda{which is equivalent to PED in gas phase molecules}, by measuring \didier{simultaneously} the momentum correlations between  \didier{the} core-level photoelectrons and  \didier{the} fragment ions.} 

\hatada{Synchrotron radiation based studies ~\cite{Williams2012,Fukuzawa2019} have shown that MFPADs averaged over the polarization angle of incident X-rays (Polarization-Averaged Molecular-Frame Photoelectron Angular Distributions: PA-MFPADs) provide information on the 
structure of molecules.}
\hatada{The beauty of polarization averaging is that the most prominent peaks produced by directly excited photoelectron waves, where the angular distribution of dipoles is parallel to the polarization vector, are smeared out, and the effect of scattering by surrounding atoms of the absorbing atoms is emphasized. Hence, PA-MFPADs reflect the  three-dimensional \didier{structure} information more clearly than MFPADs.}


\hatada{
\ueda{Several attempts aiming at} tracing molecular dissociation \didier{through} time-resolved MFPADs measurements  have already been performed~\cite{Rouzee2013, Boll2013, Nakajima2015, Minemoto2016}} 
\ueda{but without much success.
This situation is dramatically changing \didier{now} thanks to the emergence of \didier{the} European XFEL, the first high repetition-rate XFEL~\cite{Decking2020}, and \didier{of the} COLTRIMS-Reaction microscope installed there~\cite{Kastirke2020}. \didier{U}sing sequential ionization of the O $1s$ core level in the O$_2$ molecule within a single XFEL pulse, it was demonstrated that time-resolved PA-MFPADs measurements are possible with  XFEL~\cite{Kastirke2020}.  Since an XFEL-pump--XFEL-prob\didier{e} system has also been installed there~\cite{Serkez2020}, making a molecular movie recording the PA-MFPADs is no longer an intangible  dream but a tangible reality.}   
\ota{
This article is the third in a series of articles on \didier{the} \ueda{theoretical study of} PA-MFPADs 
\ueda{for dissociating} hetero-diatomic 
\ueda{dications CO$^{2+}$, which may be} measured by pump-probe 
\ueda{experiments} with \didier{the} two-color XFEL.
\ueda{
\didier{In this type of experiment, }
the first XFEL pulse removes one of the core electrons and the subsequent Auger decay creates the CO$^{2+}$ ion that starts to dissociate. Then one can probe \didier{the} variation\didier{s} of the C-O bond length by the time-resolved PA-MFPADs measurement using the second XFEL pulse as a probe.} 
In the first paper~\cite{Ota2021a} \didier{in this series}, we introduced the Full-potential Multiple Scattering \ueda{(FPMS)} theory and presented 
results for \didier{the} calculation of PA-MFPADs of CO$^{2+}$ \ueda{as a function of the C-O bond length}. 
In the second paper~\cite{ota2021b}, we proposed a fitting method 
\ueda{for \didier{the} retrieval of} the bond length 
from the EXAFS-type oscillations which appear in the backward \ueda{scattering direction} of the PA-MFPADs.}

\ueda{In this third paper, we derive an analytical formula \ueda{that describes a fringe pattern of the PA-MFPAD, which may be \didier{identified as}  {\it a flower shape}.
The formula derived \didier{can} be used \didier{to}} \didier{estimate} the bond length of \ueda{dissociating} hetero-diatomic molecules.  Because \didier{this} {\it flower shape} \didier{pattern} is generated by the interference of two electron waves emitted \didier{respectively} from the X-ray photon absorbing atom, and \didier{from} the neighboring \ueda{scattering} atom, the mechanism that constitutes this {\it flower shape} may be interpreted as an \ueda{analogue} to Young's double-slit experiment. However, the physical phenomena \didier{behind} Young's double-slit experiment and PA-MFPADs are not exactly the same.
The key difference between them is that Young's double-slit experiment represents the interference of two spherical $s$-waves, while the O $1s$ PA-MFPADs \didier{we are interested in} consists \didier{in} the interference between one $p$-wave, which is the direct wave \didier{emitted by} the atom that absorbs an X-ray photon, and \didier{the}  wave scattered \didier{by} the neighboring atom, which may be approximated by \didier{an} $s$-wave.
\ueda{\didier{Using Multiple Scattering theory}, we derive a new formula for the $p$-$s$ wave interference \didier{in} the PA-MFPADs, analogous to Young's formula for the $s$-$s$ wave interference,} 
and further examine the validity of this new formula, \didier{by} applying it to the PA-MFPADs of dissociating CO$^{2+}$ calculated \didier{within} the FPMS theory~\cite{Ota2021a}. We also discuss the issue of applying the original Young's formula to the analysis of PA-MFPADs.
}
%
%
\section{Theory}
%

\label{sec:pamfpad}
\label{sec:flowershape}
\hatada{In this section,} \ueda{we derive analytical expressions for the \hatada{\textit{flower shape}} structure 
that appears as an oscillatory structure \didier{when varying the} angles in the PA-MFPADs.}
\hatada{We first give an expression to determine the angles of \didier{the} bright and dark fringes of PA-MFPADs. Next, utilizing this \didier{result}, we derive a 
relationship between \didier{the} low and high angle fringes.}

\begin{figure*}[htb]
\includegraphics[width=\linewidth]{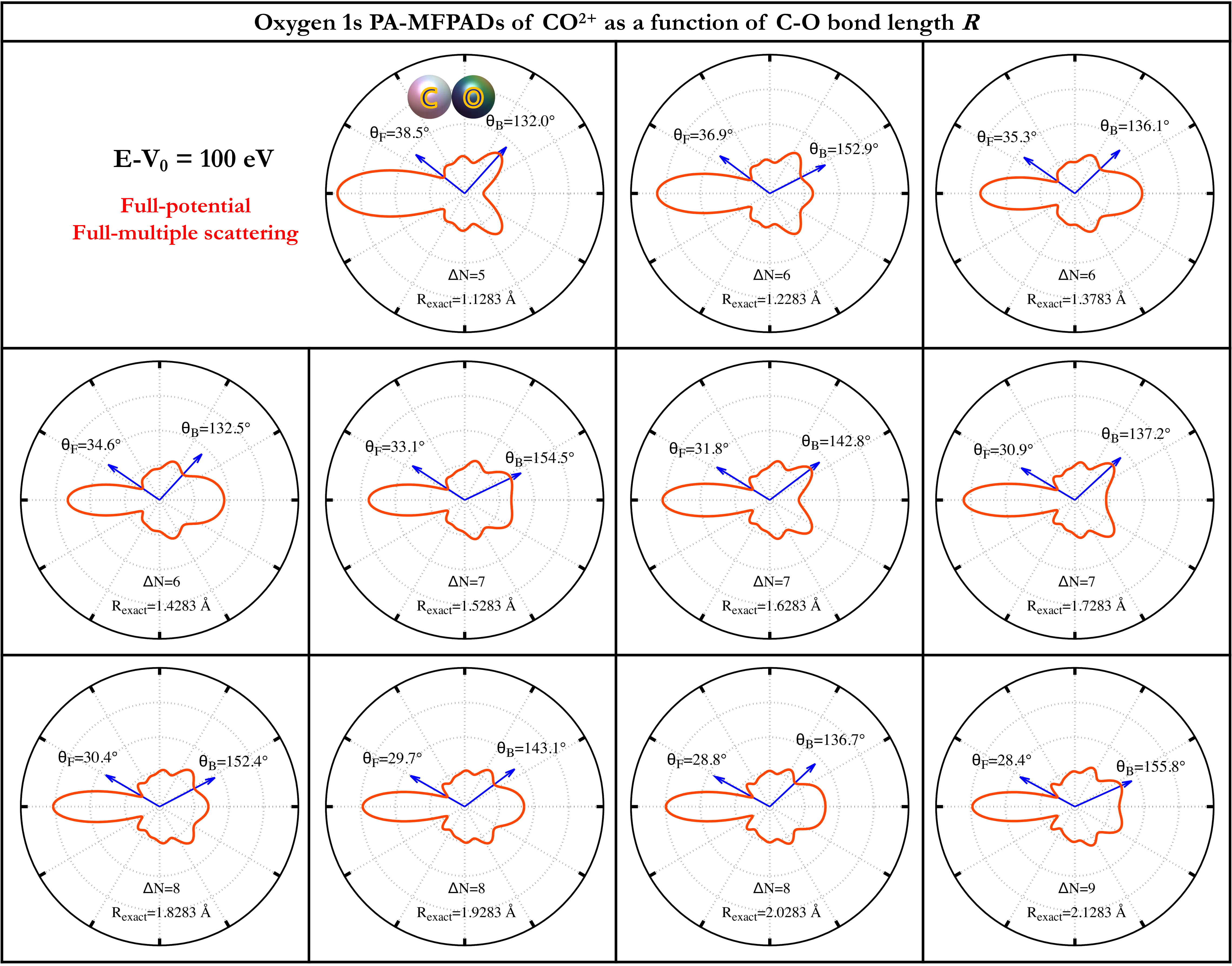}
\caption{\ota{PA-MFPADs of CO${}^{2+}$ obtained by the FPMS calculations as a function of \didier{the} C-O bond length $R$,
\didier{at a} photoelectron kinetic energy \ota{$E-V_0\,=\,100$ eV}. (See the preceding paper \cite{Ota2021a} for the details.)
Blue arrows indicate the position of the most forward valley position $\theta_F$ and the most backward peak/valley position $\theta_B$.
 $\Delta N=N_B-N_F$ is the difference \didier{in} the index numbers of the corresponding peak/valley positions.
}
\label{fig:peaks}  
}
\end{figure*}
%

Figure~
\ueda{\ref{fig:peaks} depicts} PA-MFPADs calculated \didier {as a function of} the \hatada{C-O bond length} $R$, using the FPMS theory of PA-MFPADs  
\didier{for photoelectrons of momentum $\textbf{k}$~\cite{Ota2021a}} 
\begin{eqnarray}
\left<I({\bf k})\right>_{\varepsilon}
 \equiv 
 \frac{1}{4\pi} \int d\hat{\varepsilon} \, 
 I({\bf k},\hat{\varepsilon})  \nonumber \\
 =
 \frac{1}{3}
 \left[
 I({\bf k},\hat{\bf x})
 +
 I({\bf k},\hat{\bf y})
 +
 I({\bf k},\hat{\bf z})
\right] \label{eq:pamfpads_vecave}
 \\
= \frac{8 \pi^2 \alpha \hbar \omega}{3}
 \sum_{n=-1}^{1}  \sum_{m_c}
 \Bigg|
 \sqrt{
 \frac{4\pi}{3}
 }
 \sum_{LL'}
 B_L^{\,i *} ({\bf k})
  \nonumber \\
 \hspace{0.3cm}
 \times
 C(L,1n,L_c) \int dr \,
 r^3 \,
 R_{L'L} (r;k) \,
 R_{L_c}^{\,c} (r) \,
\Bigg|^2 , \label{eq:pamfapds}
\end{eqnarray}
\hatada{where}
\begin{eqnarray}
B_L^{\,i} ({\bf k})=&
 \sum_{jL'}
 \tau_{LL'}^{ij} \,
 I_{L'}^{\,j} ({\bf k}),
\end{eqnarray}
\hatada{and}
\begin{eqnarray}
\tau &\equiv &\left( T^{-1} - G \right)^{-1}=
T \left( 1 - G T \right)^{-1}, \label{eq:tau}
\\
 I_{L}^{\,i} ({\bf k})   &\equiv& i^{\,l}
 \sqrt{\frac{k}{\pi}}\, 
 e^{i{\bf k }\cdot{\bf R}_{io}}
 \mathcal{ Y }_{L}(\hat{\bf k }).
\end{eqnarray}
The indices $i$ and $j$ 
 \didier{refer to the} scattering sites $i$ and $j$.
$\alpha$ is the fine structure constant, $\textbf{R}_{io}$ is the vector %
\didier{connecting} the origin to the center of scattering site $i$ and $\mathcal{ Y }_{L}$ is \didier {the} real spherical harmonics \didier{of} angular momentum $L=(l,m)$. In the second line of equation~\ref{eq:pamfapds}, $ C(L,1n,L_c)$ is the Gaunt coefficient \didier{composed} of real spherical harmonics and the rest is the radial integral between the local solution and \didier{the} core wave function. In equation~\ref{eq:tau}, $T$ is the transition operator and $G$ is the KKR structure factor\didier{; They are matrices} 
\hatada{ labeled with \didier{the} indices referring to} scattering sites and angular momentum~\cite{Korringa1947,Kohn1954}.

\hatada{We see that the  PA-MFPADs intensity $\left<I({\bf k})\right>_{\varepsilon}$ which is defined as the average over polarization angle of MFPADs intensity $I({\bf k},\hat{\varepsilon})$ is equivalent to the average over \didier{the} three MFPADs with \didier{the} polarization vector along \didier{the} $x$, $y$ and $z$-axes respectively.}
More details are reported in the preceding paper~\cite{Ota2021a} \hatada{\didier{in} this series}.
\didier{T}he electron charge density was calculated \yamazaki{at the RASPT2/ANO-RCC-VQZP level of theory as implemented in MOLCAS 8.2~\cite{Aquilante2016}} for \didier{the multiplet state} 
1$\sigma^{-1}$5$\sigma^{-2}$ (i.e. O 1$s^{-1}$ HOMO$^{-2}$) \didier{considered as the dominant} Auger final state~\cite{Cederbaum1991}.
\hatada{We chose \didier{for the} photoelectron energy $E-V_0=100$ eV.}
\didier{S}everal small  lobes \didier{appear} between the forward- and backward-\hatada{intensity} peaks. 
\ota{
\didier{The e}longation of the bond length $R$ moves these lobes from the backward to \didier{the} forward \didier{position} and increase\didier{s} the\didier{ir} number by one for one period of oscillation of the backward-intensity peak. 
}


\hatada{From a simple physical insight, we \didier{identify} the \textit{flower shape} structure at the intermediate angles \didier{as a signature} of \didier{the} interference between \didier{the} direct wave of \didier{the} photoelectron \didier{originating} from the oxygen atom and the  \didier{photoelectron wave scattered by} the carbon atom. This \didier{process} \didier{strongly} resembles a Young's double-slit experiment.
The following formula, the so-called Young's formula, obtained from classical wave mechanics describes \didier{the} relationship \didier{connecting} the distance $R$ \didier{between the} two slits \didier{to} \didier{the} two \didier{fringes angles}  $\theta_{\nu}$ and $\theta_{\nu+\mu}$,
\begin{eqnarray}
\label{eq:Young's_eq}
R=\frac{\, \mu\, \pi}{k\left(\cos \theta_{\nu}-\cos \theta_{\nu+\mu}\right)} \label{eq:Young} ~~\mbox{($\theta_{\nu}<\theta_{\nu+\mu}$)}~, \nu, \mu = 1, 2, ...
\end{eqnarray}
%
where the positive integers $\nu$ and $\mu$ are index numbers \didier{identifying the}  $\nu^{\rm th}$ and $(\nu+\mu)^{\rm th}$ bright or dark fringes counted from the angle zero along \didier{the} two slit axis. This \didier{formula} is conventionally used to determine the bond length of molecules for PA-MFPADs.
\yamazaki{However, \didier{it} is only valid for \didier{a} two $s$-wave interference. This is not the case of PA-MFPADs  \didier{which} originate from  the interference between \didier{the} $p$-wave of \didier{the} \ueda{O 1$s$} photoelectron ejected from the  \ueda{X-ray photon} absorbing atom mostly \didier{through an} electron dipole transition along the molecular axis, and the scattered wave from the neighbouring atom at distance $R$, \didier{which may be} approximated \didier{by an} $s$-wave.}
The purpose of this article is to derive \ota{a} \didier{corresponding} formula for \didier{this} $p$-$s$ interference \didier{within} the framework \didier{of} quantum scattering theory.
}

\begin{figure}[htb]
\includegraphics[width=\linewidth]{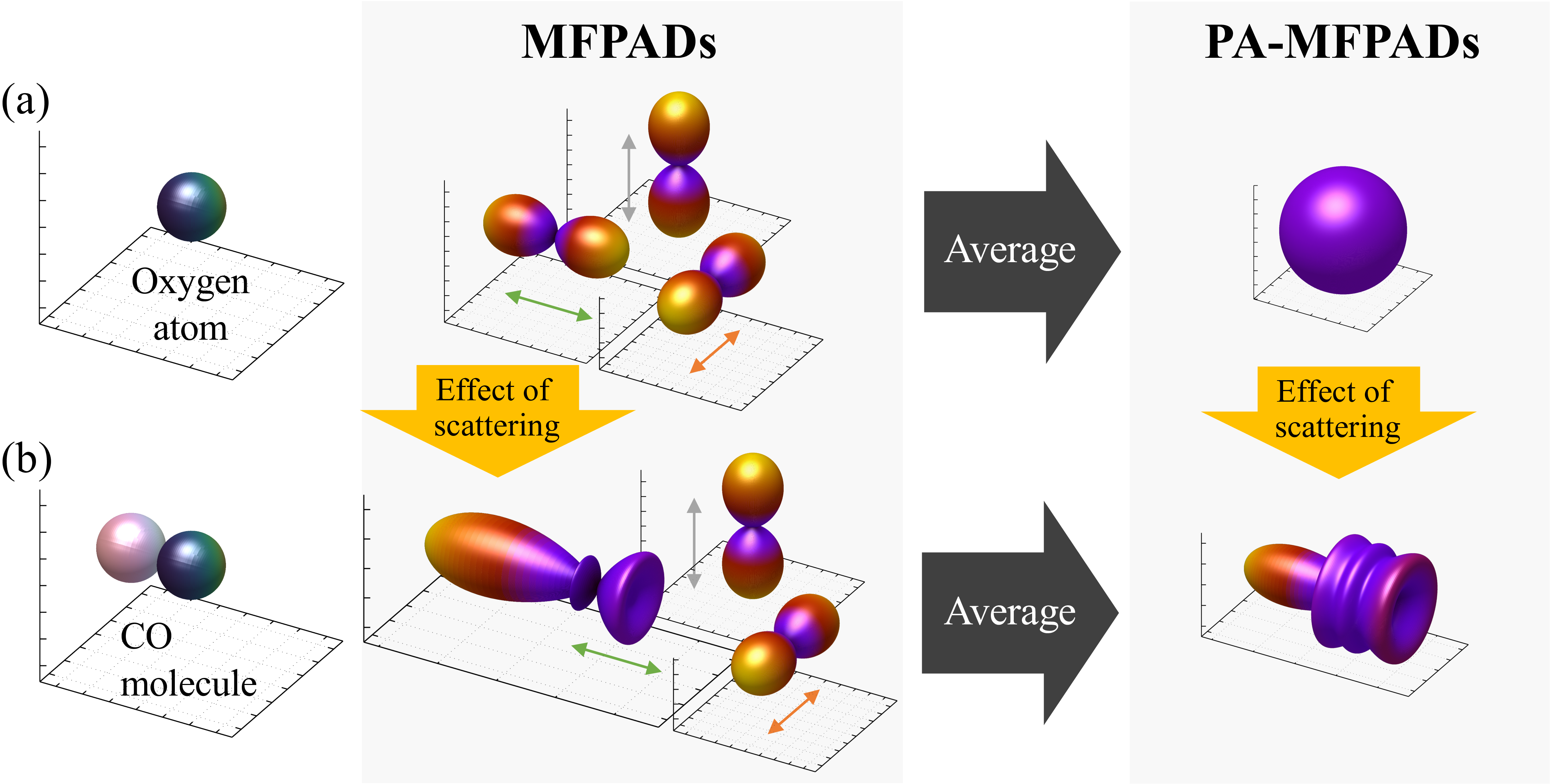}
\caption{
\label{fig:MFPADs_PA-MFPADs}
Images of O 1$s$ MFPADs and O 1$s$ PA-MFPADs of (a) an oxygen atom and (b) a CO molecule calculated \yamazaki{within the Muffin-tin, single scattering, and Plane Wave approximations.} The arrows in the figure indicate the polarization vector of the incident X-rays.
\ota{The PA-MFPADs correspond to the average of the three MFPADs excited by incident X-rays whose polarization vectors are orthogonal to each other, as shown in equation~\ref{eq:pamfpads_vecave}}.
Note that, within these approximations, \hatada{a photoelectron} excited by polarized light along the molecular axis propagate\didier{s} to the neighboring atom and \hatada{is} scattered, whereas \didier{in} MFPADs excited by polarized light perpendicular to the molecular axis, 
\didier{the photoelectron is not} 
scattered and 
\didier{is} still dipole-distributed.
}
\end{figure}
In order to obtain a relationship between $\theta$ and $R$ for PA-MFPADs, we start from the following formula for PA-MFPADs with\didier{in the single-scattering} plane wave and  Muffin-tin approximations~\cite{ota2021b},
 \begin{eqnarray}
 \left<I_{single}(k,\theta)\right>_{\varepsilon}
 =\frac{8}{3} \pi k \alpha \hbar \omega\left|t_{1}^{O} M_{00}^{1}\right|^{2}
 \nonumber \\
 \times
 \sum_{m_{p}}\Bigg\{ \left|
 \mathcal{Y}_{1 m_{p}}(-\hat{\mathbf{k}})\right|^{2} 
 \nonumber \\
 \hspace{1cm}
 +\frac{2 \Re\left(e^{i k R\left(1-\cos\theta\right)} f^{C}\left(k, \theta \right)\right)}{R} 
 \mathcal{Y}_{1 m_{p}}(-\hat{\mathbf{k}}) 
 \mathcal{Y}_{1 m_{p}}\left(-\hat{\mathbf{R}}\right)
 \nonumber \\
\hspace{1cm} +\frac{\left|f^{C}\left(k, \theta \right)\right|^{2}}{R^{2}}
\left|\mathcal{Y}_{1 m_{p}}\left(-\hat{\mathbf{R}}\right)\right|^{2}\Bigg\},
\label{eq:single_scat_mp}
 \end{eqnarray}
 \ota{
  where 
$f^{\,C}\,(k,\theta)
 \equiv
 - 4\pi
   \sum_{L}
   T_{l}^{C} \,
   \mathcal{Y}_L(\hat{\bf k}) \,
   \mathcal{Y}_L(\hat{\bf r})
$
is \didier{the} scattering amplitude
and the superscripts denote the scattering atom,  i.e. $C$ for carbon and $O$ for oxygen.
 \didier{We remind} that, \didier{as} the CO$^{2+}$ 
 is arranged to have its molecular axis parallel to the $z$-axis, $\mathcal{Y}_{1 \pm 1}\left(-\hat{\mathbf{R}}\right)=0$ and $\mathcal{Y}_{10}\left(-\hat{\mathbf{R}}\right)=-1$. Thus, the second and third terms in equation \ref{eq:single_scat_mp} are due only to the component $m_p=0$.}
\yamazaki{In \didier{this} physical picture, only O $1s$ photoelectron\didier{s} excited \ueda{ by X-rays} polarized parallel to the molecular axis propagate to \didier{the} neighboring atom and are scattered, whereas the one\didier{s} excited by  \ueda{X-rays with \didier{a} polarization perpendicular to the molecular axis} are not scattered, as \didier{explicited} in \didier{the} \hatada{MFPADs in } figure~\ref{fig:MFPADs_PA-MFPADs}(b).
 }
 
 \didier{Performing} the summation \didier{over} $m_p$, we obtain
 \begin{eqnarray}
& \left<I_{single}(k,\theta)\right>_{\varepsilon}
 =
  2 k \alpha \hbar \omega
 \left|
  T_{1}^{\,O} \,
  M_{00}^{1}
  \, \right|^2
&
 \nonumber \\
& \times
  \Bigg\{
   1
   +
   \frac{ 2
   \Re \left(
	\,e^{ikR_{}(1-\cos \theta ) }
	f^{\,C}\,(k,\theta) \,
       \right)
   }{\,R_{} } \,
   \cos \theta
   +	 
   \frac{\,\left|f^{\,C}\,(k,\theta)\right|^2 }{\,R_{}^2 } \,
  \Bigg\}.& 
  \label{eq:ave-single-co}
 \end{eqnarray}
\hatada{The summation corresponds to the sum of \didier{the} MFPADs excited by three orthogonal linearly polarized lights shown in the center of figure~\ref{fig:MFPADs_PA-MFPADs}(b) and the result of equation~\ref{eq:ave-single-co} is the PA-MFPADs shown \didier{in the} right hand \didier{side} in figure~\ref{fig:MFPADs_PA-MFPADs}(b).}
\ota{
The first term corresponds to \didier{the} direct photoemission \didier{process} 
\didier{where no scattering occurs after emission by} the photo-absorbing atom\hatada{. 
This term corresponds to the atomic PA-MFPADs in figure~\ref{fig:MFPADs_PA-MFPADs}(a), and indeed it does not depend on the angles.
T}he third term corresponds to the intensity of a wave singl\didier{y} scattered by the neighboring carbon atom, and the second term corresponds to the interference between the direct wave and the singl\didier{y} scattered wave. This second term 
\didier{gives rise to} the fringes \didier{observed} at intermediate angles\hatada{, which \didier{can be identified to the}  wrinkles in \didier{the} PA-MFPADs in figure~\ref{fig:MFPADs_PA-MFPADs}(b)}, \didier{and which we refer to as a} "{\it flower shape}" \didier{pattern}.} 

\hatada{
We define \didier{now} the following function which is responsible 
\didier{for} the \textit{flower shape} in the second term in equation~\ref{eq:ave-single-co},
}
\begin{eqnarray}
F(\theta) \equiv \cos \,(\,k R\,(\,1-\cos \theta\,)+\phi(\theta)\,) \cos \theta,
\label{eq:F_theta}
\end{eqnarray}
\hatada{where $\phi(\theta)\equiv \phi(k, \theta)$ is the phase function of the scattering amplitude,
$f^{c}(k, \theta)=\left|f^{c}(k, \theta)\right| \exp{[\,i \, \phi(k, \theta)]}$. \didier{The modulus function}  $|f^{c}(k, \theta)|$ is a smooth function 
\didier{of} the angle $\theta$ \didier{when} compared to the rest in the second term in equation~\ref{eq:ave-single-co}, so that we \didier{can} neglect the $\theta$ dependence \didier{in} $|f^{c}(k, \theta)|$ 
for the study of \didier{the} {\it flower shape}.
We hereafter omit the \hatada{argument} $k$ since we focus on PA-MFPADs at \hatada{a} \didier{given} energy.}
\hatada{The zeros \didier{in the} derivative of this function with respect to $\theta$ correspond to the positions of fringes. Thus, the equation to be solved is}\hatada{
\begin{eqnarray}
F'(\theta) &=& -\sin\theta \, \Bigg[
\cos \,(\,k R\,(\,1-\cos \theta\,)+\phi(\theta)\,) \nonumber \\
&&\hspace{-1cm}+ \sin \,(\,k R\,(\,1-\cos \theta\,)+\phi(\theta)\,)\,
\left(
kR-\frac{d\phi(\theta)}{d\cos\theta}
\right)\cos\theta
\Bigg]=0. \nonumber
\\
\label{eq:eqtobesolved}
\end{eqnarray}
The zeros at $\theta=0$ and $\pi$ from the $\sin\theta$ are 
\didier{of no interest to} us since \didier{there}, we can not distinguish the contribution 
\didier{to the} interference\didier{s} from \didier{that of the} large forward- and backward-intensities. 
We 
\didier{therefore look for the }zeros inside the  brackets \didier{ $[ \, ]$} and the equation to be solved 
\didier{becomes}}
\begin{eqnarray}
k R\,(\,1-\cos \theta\,)+\phi(\theta)\, +
\arctan
\left(
\frac{1}{
\left(kR-\frac{d\phi(\theta)}{d\cos\theta}\right)\,\cos\theta
}
\right)
\nonumber
=n\pi, \label{eq:eqtobesolved_flower} \\
\end{eqnarray}
where $n$ is an integer.
This equation can not be solved exactly and \didier{therefore} we introduce some 
\didier{simplifications in order} to obtain approximate solutions. 

\ota{Here, we assume that $\left(kR-\frac{d\phi(\theta)}{d\cos\theta}\right)>0$.}
\hatada{In order to make the integer $n$ correspond to the numbering of the fringe,}
\ota{ the LHS of equation \ref{eq:eqtobesolved_flower} needs to be a function that increases as $\theta$ increases from 0 to $\pi$. On the other hand, the value of the arc-tangent function jumps by $\pi$ at $\theta=\pi/2$ where the sign of the argument changes. \didier{So, in order} to avoid this jump and map the integer $n$ to the \didier{identification number} of the peaks, i.e. the fringes, we add $\pi$ to the LHS when the argument of arc-tangent is negative, i.e. in \didier{the} backward region ($\pi/2<\theta<\pi$).
}

\didier{We note respectively $\theta_F$ and $\theta_B$ the peak positions close to $\theta=0$ and $\theta=\pi$. }
\didier{Then, e}quation \ref{eq:eqtobesolved_flower} becomes
\begin{eqnarray}
k R\,(\,1-\cos \theta_F\,)
+\phi(\theta_F)\, \nonumber \\
+\arctan
\left(
\frac{1}{
\left(kR-\left.\frac{d\phi(\theta)}{d\cos\theta}\right|_{\theta=\theta_F}\right)\,\cos\theta_F
}
\right)
=N_F\pi, \\
k R\,(\,1-\cos \theta_B\,)
+\phi(\theta_B)\, \nonumber \\
+\arctan
\left(
\frac{1}{
\left(kR-\left.\frac{d\phi(\theta)}{d\cos\theta}\right|_{\theta=\theta_B}\right)\,\cos\theta_B
}
\right)
+\pi
=N_B\pi,
\end{eqnarray}
where 
the integers $N_F$ and $N_B$ are 
\didier{the} number of peaks and valleys of PA-MFPADs at $\theta_F$ and $\theta_B$ counted from the forward to the backward direction \didier{respectively}. \hatada{In figure~\ref{fig:peaks}, these angles for \didier{each} PA-MFPADs \didier{are displayed} \hatada{by the arrows}.}

\didier{A}ssuming that 
$\left|
\left(kR-\frac{d\phi(\theta)}{d\cos\theta}\right)\cos\theta\right|
\gg 1$ above 
\didier{medium values of the photoelectron energy} ($\gtrsim 100$ eV),
the arc-tangent function
becomes negligibly small. Then the \didier{previous} set of equations become\didier{s},
\begin{eqnarray}
k R\,(\,1-\cos \theta_F\,)+\phi(\theta_F)\,
\sim N_F\pi, \label{eq:forward} \\
k R\,(\,1-\cos \theta_B\,)+\phi(\theta_B)\,
+\pi
\sim N_B\pi
= (N_F+\Delta N)\pi, \label{eq:backward} 
\end{eqnarray}
where $\Delta N$ is 
\didier{the} difference between $N_F$ and $N_B$. 
This approximation works well when $\theta_{F}$ is close to 0, and
$\theta_{B}$ 
to $\pi$.
\didier{T}aking \didier{the} difference between equations~\ref{eq:forward} and \ref{eq:backward},
\hatada{we obtain,}
\begin{eqnarray}
k R\,(\,\cos \theta_F-\cos \theta_B\,)+\phi(\theta_B)\,
-\phi(\theta_F)\,
\sim (\Delta N-1)\pi \label{eq:forward_backward}.
\end{eqnarray}

The phase function $\phi(\theta)$ may further be approximated \didier{by a} linear function of $\cos\theta$\didier{:} $\phi(\theta)\sim
A+B\cos\theta$. 
We determine the coefficients $A$ and $B$ at the equilibrium bond length $R_{\rm eq}$, \didier{and obtain}
\begin{eqnarray}
%
A=
\frac{
\phi(\theta_B^{\rm eq})\cos\theta_F^{\rm eq}-\phi(\theta_F^{\rm eq})\cos\theta_B^{\rm eq}
}{\cos\theta_F^{\rm eq}-\cos\theta_B^{\rm eq}},
\label{eq:cos_phaseA}
\\
B=
\frac{\phi(\theta_F^{\rm eq})-\phi(\theta_B^{\rm eq})}{\cos\theta_F^{\rm eq}-\cos\theta_B^{\rm eq}},
\label{eq:cos_phase}
\end{eqnarray}
where $\theta_F^{\rm eq}$ and $\theta_B^{\rm eq}$ are \didier{respectively} $\theta_F$ and $\theta_B$ at the equilibrium bond length.

\begin{figure}[htb]
\includegraphics[width=\linewidth]{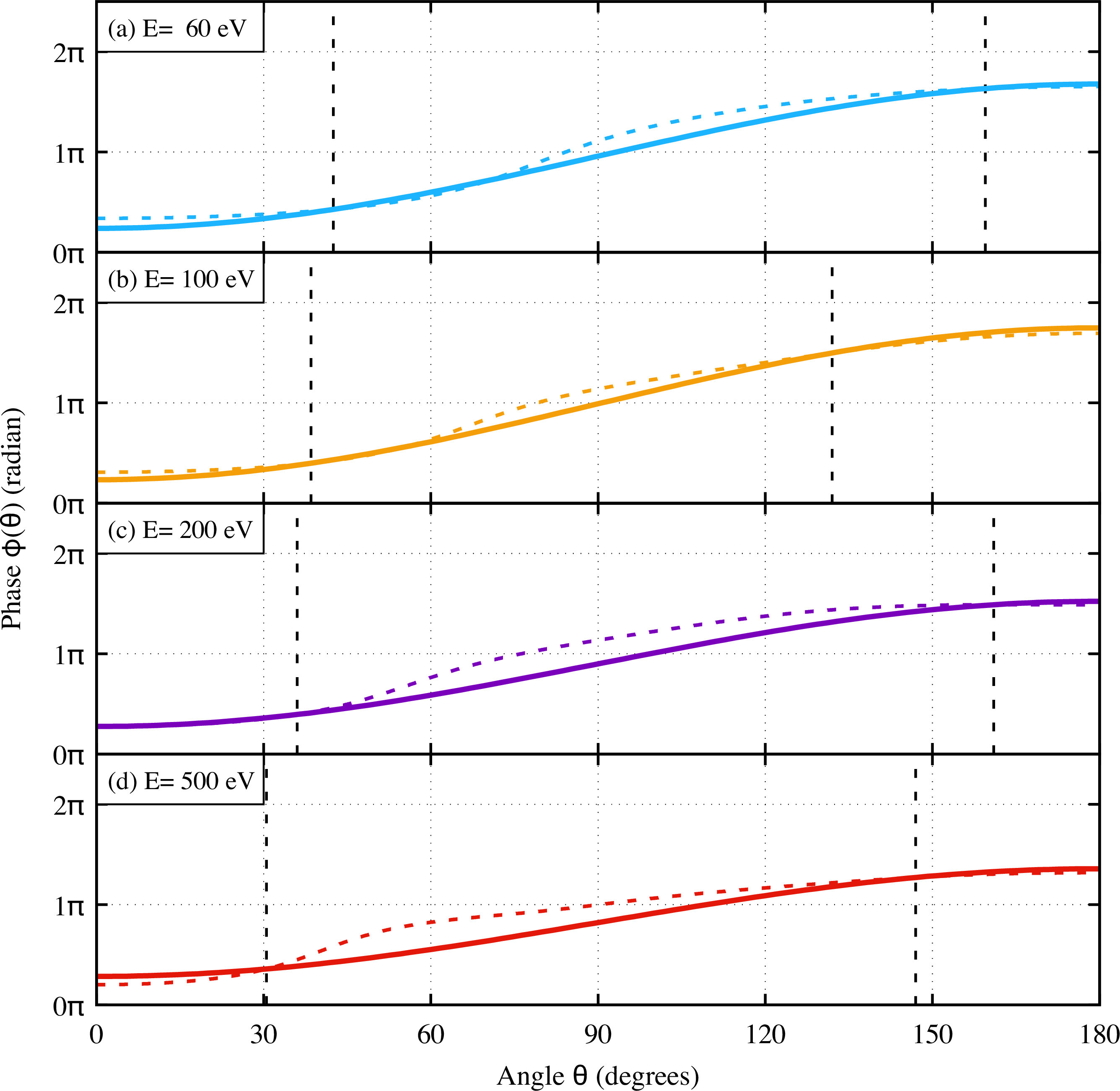}
\caption{
The dashed curves represent
the phase $\phi(k,\theta)$ of the scattering amplitude of the C atom in the CO molecule with a ground state bond length $R_{\rm eq}$ numerically calculated using MsSpec~\cite{Sebilleau2011}. \revise{
The solid curves are the approximate phase functions $A+B\cos\theta$ where the coefficients $A$ and $B$ are defined in equations~\ref{eq:cos_phaseA} and ~\ref{eq:cos_phase}, respectively.} 
The photoelectron kinetic energies are (a) 60 eV,
(b) 100 eV,
(c) 200 eV,
and (d) 500 eV, respectively.
The black vertical dashed lines indicate the angles of the most forward valley $\theta_F^{\rm eq}$ and of the most backward peak/valley $\theta_B^{\rm eq}$
\revise{
: (a) $\theta_F^{\rm eq}=42.5^{\circ}$ and 
$\theta_B^{\rm eq}=159.5^{\circ}$, 
(b) $\theta_F^{\rm eq}=38.5^{\circ}$ and 
$\theta_B^{\rm eq}=132.0^{\circ}$, 
(c) $\theta_F^{\rm eq}=36.0^{\circ}$ and 
$\theta_B^{\rm eq}=161.0^{\circ}$, 
(d) $\theta_F^{\rm eq}=30.5^{\circ}$ and
$\theta_B^{\rm eq}=147.0^{\circ}$. 
}
\label{fig:phase}
}
\end{figure}

\ota{
Figure \ref{fig:phase} shows that 
the phase $\phi(\theta)$ and the 
\didier{approximate} phase function $(A+B\cos\theta)$ 
\didier{coincide} at $\theta_F^{\rm eq}$ and $\theta_B^{\rm eq}$.
According to figure \ref{fig:peaks}, the most forward valley position $\theta_F$ moves \didier{further} forward (i.e. $\theta_F<\theta_F^{\rm eq}$) and $\theta_B$ appears around $\theta_B^{\rm eq}$ \ota{when elongating the bond length $R$}.
\revise{
The phase function $\phi(\theta)$ depends on the scattering potential of scattering site of carbon, and the change of the potential with the bond elongation is negligible in the high energy regime.
}
Based on \revise{these facts} and the agreement between the numerically calculated phase and \didier{the} approximate phase function shown in figure \ref{fig:phase}, we expect this approximation of the phase \didier{to be} still valid after the elongation of the bond length $R$. }
\\
%

\didier{Applying} the approximation above to equation~\ref{eq:forward_backward}, 
we obtain \didier{finally} the equation,
\begin{eqnarray}
R\,
\sim \frac{(\Delta N-1)\pi}{k(\,\cos \theta_F-\cos \theta_B\,)}
+\beta,\label{eq:new_eq}
\end{eqnarray}
where $\beta\equiv B/{k}$, 
 \didier{ depends} on $k$ and \didier{is} \hatada{not sensitive to $R$.} 
This equation 
\didier{describes} the relationship between the bond length $R$ and the peak or valley positions of \didier{the} PA-MFPADs. \didier{It} has a form similar to Young's formula in equation \ref{eq:Young}, but the numerator 
\didier{differs} by $\pi$ and 
\didier{a} constant shift $\beta$ appears, due to the \hatada{angle dependency of the} phase $\phi(\theta)$. In the next section, we will give \didier{a} physical interpretation \didier{of this} difference between equation \ref{eq:new_eq} and \didier{the} ordinary Young's formula.
\section{Results and Discussion}
\subsection{\ueda{
Extracting bond length information from PA-MFPADs by using our new formula for $p$-$s$ interference
}}
Here, 
\didier{we} consider 
\ota{an} 
experimental procedure to determine the evolution of the bond length of diatomic molecules.
\ota{
Since the parameter $\beta$ is almost independent of the bond length, once $\beta$ is obtained from the PA-MFPADs with known bond length (e.g., equilibrium state or pre-dissociation structure), we can extract the bond length from the PA-MFPADs obtained with subsequent time evolution.
Using this strategy, the bond length $R$ can be practically evaluated as a function of the pump-probe delay time $t$ using  equation \ref{eq:new_eq}.
}

\begin{figure}[htb]
\includegraphics[width=1.0\linewidth]{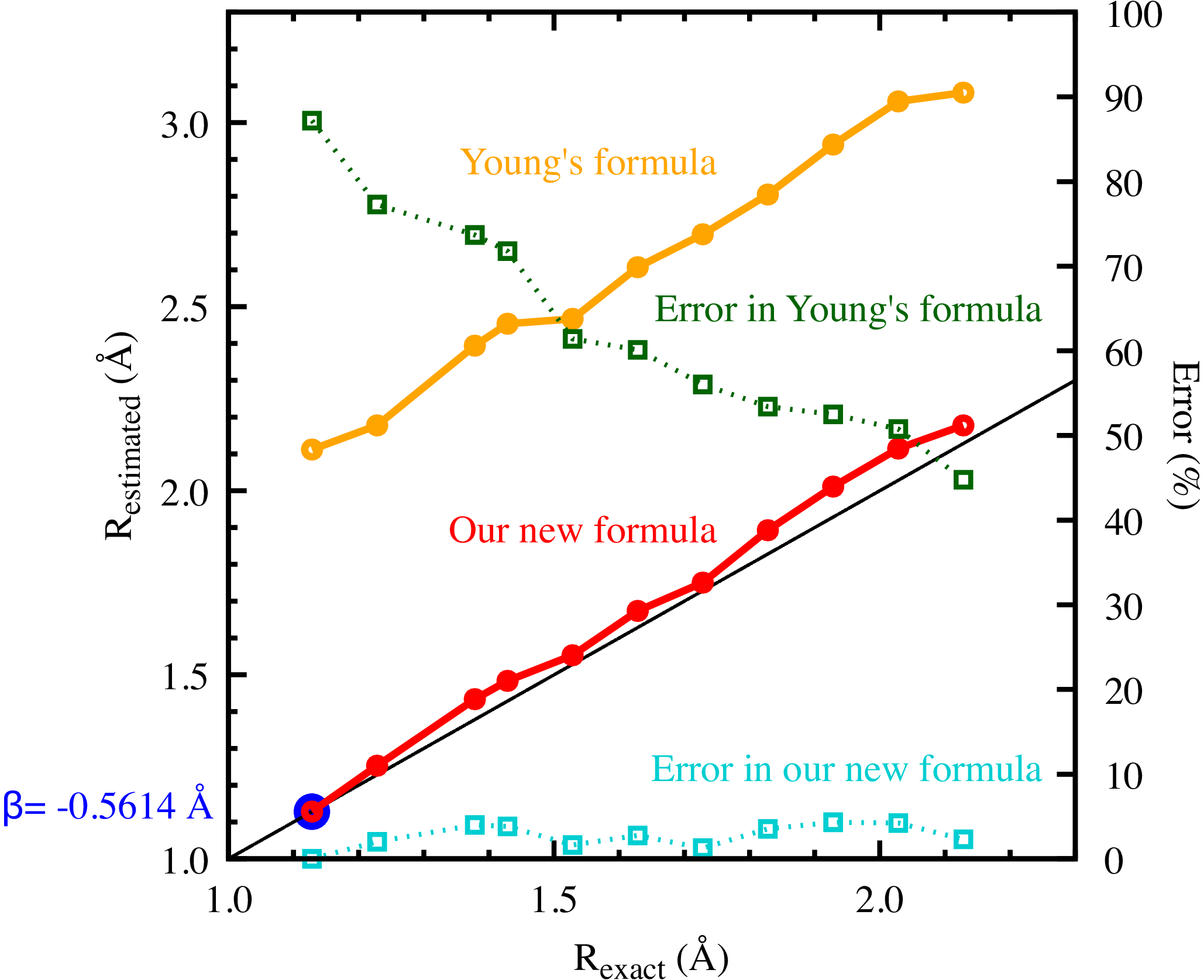}
\caption{\label{fig:R_estimated}
\ota{\didier{The r}ed and orange lines %
\didier{represent the} C-O bond length estimated \didier{respectively} from \didier{the} O 1$s$ PA-MFPADs of CO$^{2+}$ \didier{(}shown in figure~\ref{fig:peaks}\didier{)} by using our new formula in equation \ref{eq:new_eq} and \didier{from the} ordinary Young's formula in equation~\ref{eq:Young's_eq}.
The parameter $\beta$ was determined as $-0.5614$ \AA~ at $R_{\rm exact}=1.1283$ {\AA} (equilibrium bond length, $R_{\rm eq}$)\didier{; It is} marked \didier{by a} blue dot.
\didier{The s}ky blue and \didier{the} green dotted lines are \didier{respectively the } errors 
\didier{in} the estimation using our new formula in equation \ref{eq:new_eq} and \didier{using the} ordinary Young's formula in equation~\ref{eq:Young's_eq}.
The errors \hatada{in} our new formula are less than 5\% for all the calculated points.
}}
\end{figure}
\par
\ota{
Figure~\ref{fig:R_estimated} shows the estimated bond lengths $R_{\rm estimated}$ against the exact bond lengths $R_{\rm exact}$.
The estimation\didier{s were} performed by using\didier{: (i)} our new formula in equation~\ref{eq:new_eq} and \didier{(ii) the} ordinary Young's formula in equation~\ref{eq:Young's_eq}
\revise{where $\mu=\Delta N$, $\theta_{\nu}=\theta_{F}$, and $\theta_{\nu+\mu}=\theta_{B}$ were chosen,}
%
%
for the set of PA-MFPADs shown in figure \ref{fig:peaks}, 
\didier{which} were calculated with \didier{the} FPMS method \didier{at the} photoelectron energy $E-V_0\,=\,100$ eV~($k=2.711~{\rm a.u.}^{-1}$).
The parameter $\beta$ was evaluated as -0.5614 {\AA} so as to make $R_{\rm exact}$ and $R_{\rm estimated}$ agree at the equilibrium bond length $R_{\rm eq}=1.1283$ {\AA}.
The bond lengths were estimated with less than \didier{a} 5 \% 
error by using our new formula, whereas the conventional Young's formula gives %
\didier{an estimated error of} more than 50 \%. 
From these results we confirm that our new formula \didier{(}equation~\ref{eq:new_eq}\didier{)} for $p$-$s$ wave interference 
works very well for \didier{a} wide range of bond lengths, \didier{and therefore can be considered to be more or less universal}.}
\revise{
One may notice that, in figure \ref{fig:R_estimated}, if the Young's formula is shifted down by $\beta'$ to cross the equilibrium bond length $R_{\rm exact}$ (a large blue dot):
\begin{equation}
R\sim\frac{\Delta N\pi}{k(\cos\theta_F-\cos\theta_B)}+\beta'
\label{eq:Young_shift}
\end{equation}
then equation \ref{eq:Young_shift} also fits to the black line well, 
though there is no physical background of shifting the Young's formula by $\beta'$.
The reason why this artificial shift also works may be understood because, 
from equations \ref{eq:new_eq} and \ref{eq:Young_shift}, 
the shift may be given by 
$\beta'=-\pi/{k(\cos\theta_{F}^{\rm eq}-\cos\theta_{B}^{\rm eq})}+\beta$, 
with $\theta_F^{\rm eq}$ and $\theta_B^{\rm eq}$ being at the equilibrium bond length $R_{\rm eq}$, and 
$(\cos\theta_F- \cos\theta_B)$ are nearly constant (${\pi}/{k(\cos\theta_F -\cos\theta_B) \sim 0.7}$) as long as $\theta_F$ and $\theta_B$ are selected to be the closest to $\pi$ and 0, respectively, as shown in figure 1. 
}

{

The estimation \didier{obtained with} our new formula, shown in figure \ref{fig:R_estimated} 
\didier{as a} red line, exhibits a small deviation from \didier{the black} $R_{\rm exact}$,  \didier{which oscillates} as a function of $R$.
This deviation is due to the 
\didier{linear fitting} of \didier{the} phase $\phi(\theta)$ 
by a cosine function, as shown in figure \ref{fig:R_estimated}.
This fitting is done \didier{so that } 
\didier{both} functions \didier{match} at $\theta_F^{\rm eq}$ and $\theta_B^{\rm eq}$.
As the bond length elongates, the positions of $\theta_F$ and $\theta_B$ change and move away from $\theta_F^{\rm eq}$ and $\theta_B^{\rm eq}$, 
so that the errors 
\didier{in our} approximation increase.
While $\theta_F$ keeps moving away from $\theta_F^{\rm eq}$ \didier{as the} bond length \didier{increases}, the difference between $\theta_B$ and $\theta_B^{\rm eq}$ oscillates as a function of \didier{the} bond length $R$ due to the appearance of a new peak or valley. 
\didier{T}his results in 
oscillation\didier{s} 
\didier{in} the estimat\didier{ed} error.

There is 
an \didier{additional} error due to \didier{the} truncation of the Multiple Scattering series 
\didier{at first order}.
Considering that the \didier{effect of} higher order scattering 
 is not negligible for shorter bond lengths, and that our new formula was derived on the basis of the single scattering approximation, 
one would expect 
longer bond lengths 
\didier{to} give better estimations.
Contrary\didier{ly} to this expectation, figure \ref{fig:R_estimated} shows that the agreement between $R_{\rm estimated}$ and $R_{\rm exact}$ becomes worse as the bond length \didier{increases}.
This result is due to the fact that the parameter $\beta$ was evaluated at the shortest bond length which is implicitly affected by higher order scattering effect\didier{s}.

%
}

\subsection{
Using simple wave mechanics to interpret the origin of the difference between Young's formula and our new formula}\label{sec:interpretation}
\begin{figure}[htb]
\includegraphics[width=1.0\linewidth]{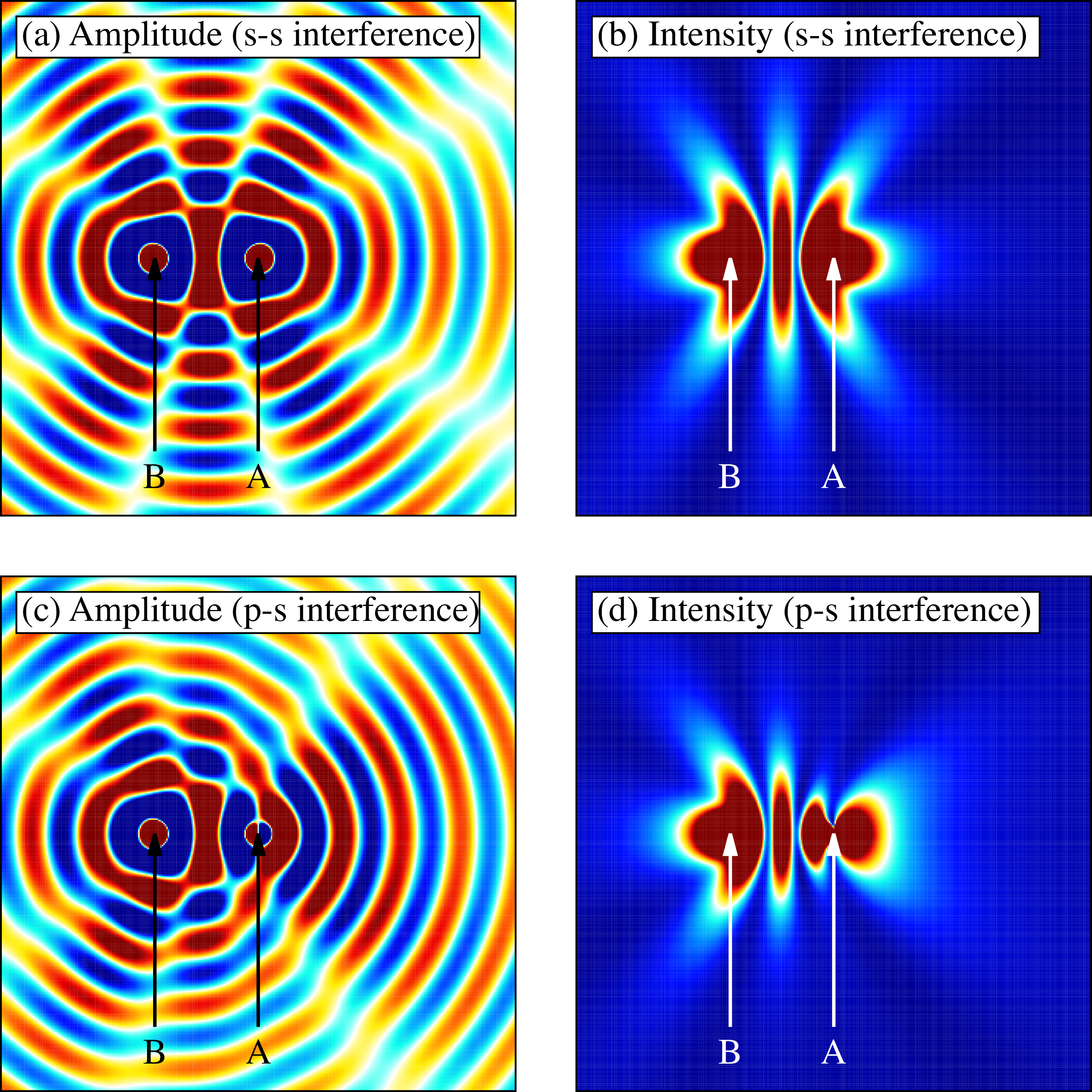}
\caption{
\label{fig:without_propagation}
Images of \didier{the} interference patterns of 
waves emitted from source points A and B.
Either \didier{an} $s$-wave \didier{(}$e^{ikr}/r$\didier{)} or \didier{a} $p$-wave \didier{(}$\cos \theta \, e^{ikr}/r$\didier{)} \didier{is} emitted from source point A, and \didier{an} $s$-wave from source point B.
(a) Real part of \didier{the} sum of the amplitudes of $s$-waves centered at points A and B,
(b) \didier{the corresponding} intensity,
(c) real part of \didier{the} summation of \didier{the} amplitudes of \didier{a} $p$-wave centered at point A and \didier{an} $s$-wave centered at point B,
and (d) \didier{the corresponding} intensity.
The bond length $R$ between A and B is set to $R=2.1283$~{\AA} $ = 4.022$ a.u. and the wavenumber \didier{to} $k=2.711$ a.u.$^{-1}$.
}
\end{figure}

\begin{figure}[htb]
\centering
\includegraphics[width=0.9\linewidth]{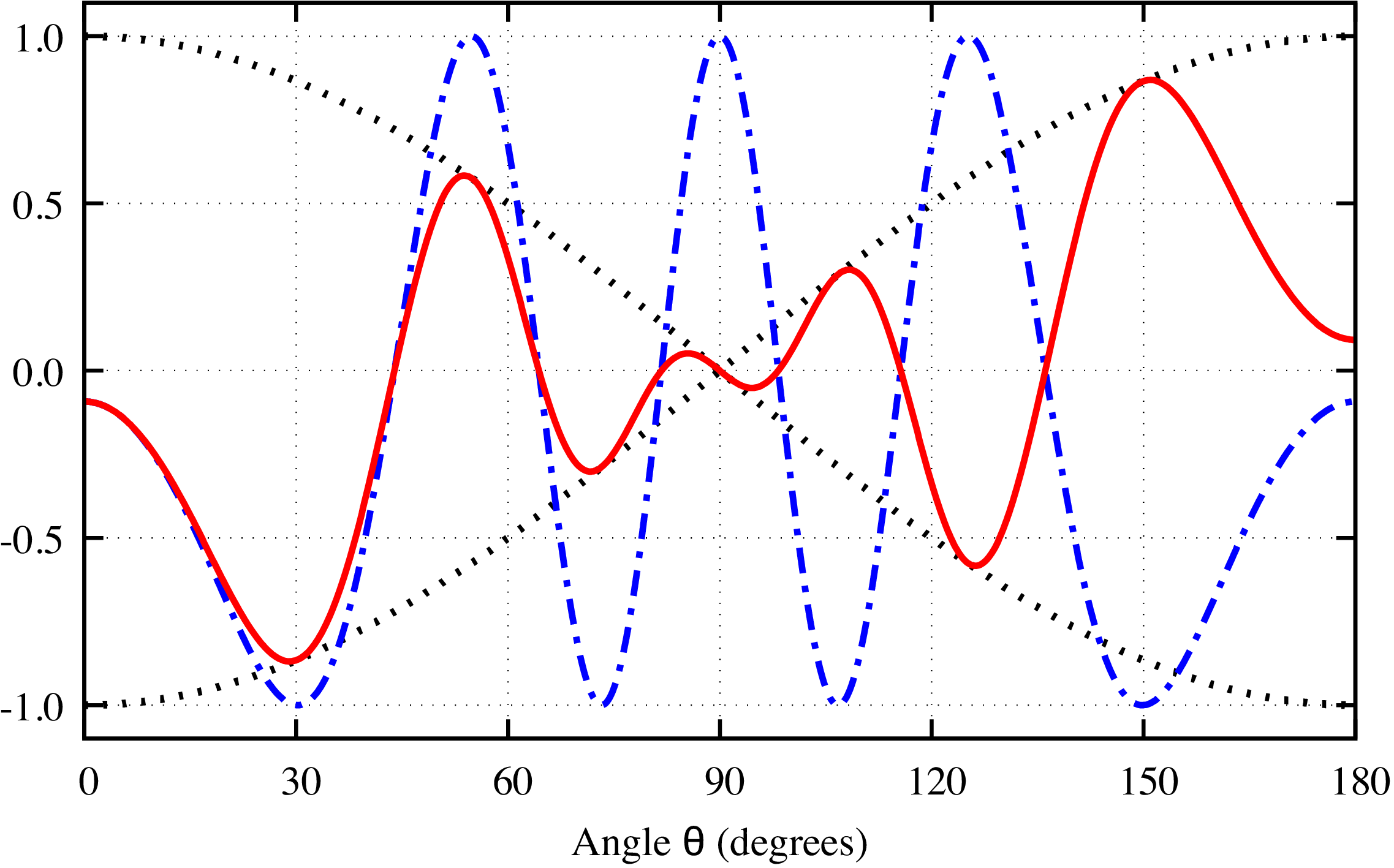}
\caption{\label{fig:cos-cos}
Comparison between $\cos (kR(\cos\theta))$ which corresponds to \didier{a} $s$-$s$-wave interference (blue \ota{dashed} curve), and $\cos (kR(\cos\theta))\cos\theta$ which corresponds to \didier{a} $p$-$s$-wave interference (red \ota{solid} curve) for $R=2.1283$~{\AA}$ = 4.022$ a.u. and $k=2.711$ a.u.$^{-1}$. The black dashed lines show the envelope of the $p$-wave, namely $\cos \theta$. The $s$-$s$ interference curve \didier{exhibits} nine \didier{zero derivative} points , while the $p$-$s$ interference curve has ten points, namely one \didier{extra} point.
}
\end{figure}

\begin{figure}[htb]
\includegraphics[width=1.0\linewidth]{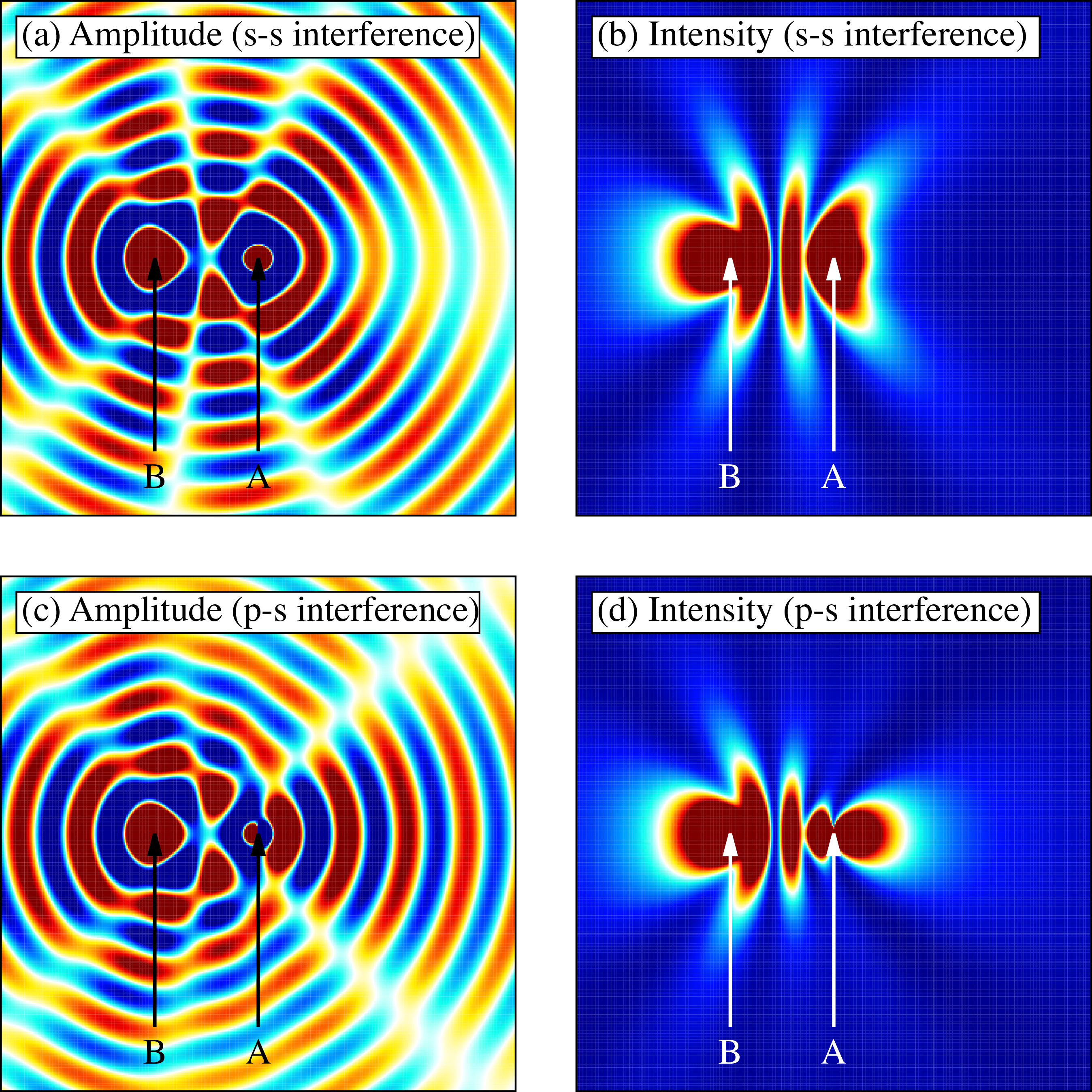}
\caption{
\label{fig:with_propagation}
Images of \didier{the} interference patterns of 
waves emitted \didier{from} source points A and B.
Either \didier{an} $s$-wave \didier{(}$e^{ikr}/r$\didier{)} or \didier{a} $p$-wave \didier{(}$\cos \theta \, e^{ikr}/r$\didier{)} \didier{is} emitted from point A, then propagate\didier{s} to point B \didier{as} a plane wave $e^{i{\bf k}'\cdot{\bf R}}$ and \didier{is} scattered\didier{; Finally} the scattered $s$-wave is emitted from source point B.
(a) \didier{r}eal part of \didier{the} sum of the amplitudes of $s$-waves centered at points A and B,
(b) \didier{the corresponding} intensity,
(c) real part of \didier{the} summation of \didier{the} amplitudes of \didier{a} $p$-wave centered at point A and \didier{an} $s$-wave centered at point B,
and (d) \didier{the corresponding} intensity.
The distance $R$ between A and B is set to $R=2.1283$~{\AA}$ = 4.022$ a.u. and the wavenumber \didier{to} $k=2.711$ a.u.$^{-1}$.
}
\end{figure}

\begin{figure}[htb]
\centering
\includegraphics[width=0.9\linewidth]{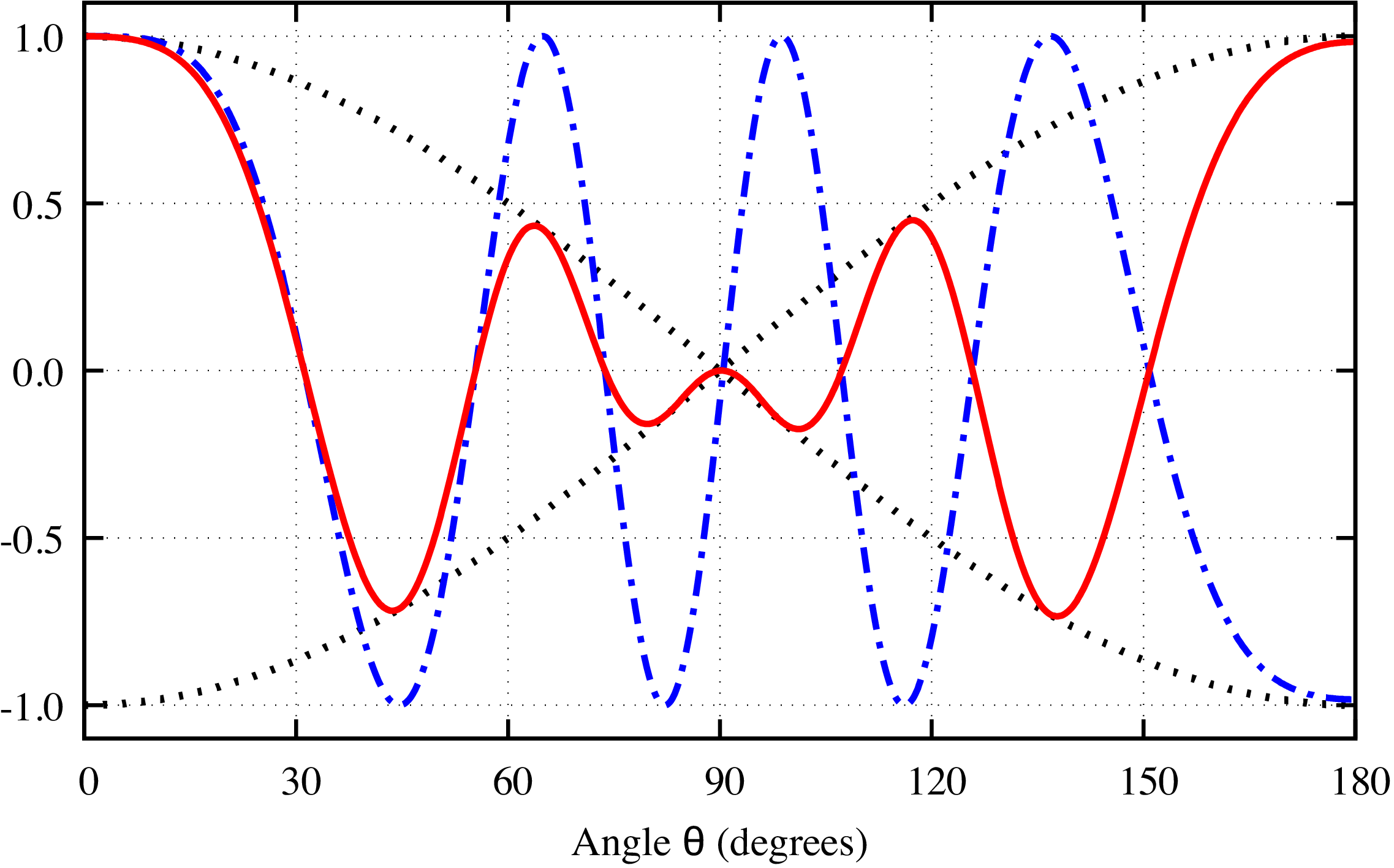}
\caption{\label{fig:cos-cos-1}
Comparison between $\cos(kR(1-\cos\theta))$ which corresponds to \didier{an} $s$-$s$-wave interference \didier{taking} propagation \didier{into account} (blue \ota{dashed} curve), and $\cos(kR(1-\cos\theta))\cos\theta$ which corresponds to \didier{a} $p$-$s$-wave interference \didier{incorporating also the} propagation (red \ota{solid} curve) for
$R=2.1283$~{\AA}$=4.022$ a.u. and $k=2.711$ a.u.$^{-1}$. The black dashed lines show the envelope of the $p$-wave, namely $\cos \theta$. The $s$-$s$ interference curve has nine \didier{zero derivative }points, while the $p$-$s$ interference curve has ten, namely one \didier{extra} point.
}
\end{figure}

From \didier{a physical} point of view, the difference between Young’s double-slit experiment and PA-MFPADs is that the {\it flower shape} \didier{pattern in} PA-MFPADs is mainly composed of interference\didier{s} between a $p$-wave (directly excited dipole wave \didier{originating} from the atom absorbing \didier{the} X-ray) and an $s$-wave (wave singly scattered from the neighboring atom). On the other hand, Young’s double-slit experiment is composed of interferences between an $s$-wave and another $s$-wave. Figures \ref{fig:without_propagation}-\ref{fig:cos-cos-1} show the $s$-$s$ and $p$-$s$ interference\didier{s} using a very simple model in which the $s$-wave is $e^{ikr}/r$ and the $p$-wave is $\cos\theta~e^{ikr}/r$.
%
The difference between figures \ref{fig:without_propagation}-\ref{fig:cos-cos} and figures \ref{fig:with_propagation}-\ref{fig:cos-cos-1} is %
\didier{that in the latter, we have incorporated a plane wave propagation $e^{i{\bf k'}\cdot{\bf R}}$, where ${\bf k'}\equiv k {\hat{\bf R}}$,} 
\didier{for} the photoelectron wave \didier{going from }point A to point B. 
In figure \ref{fig:without_propagation}, either \didier{an} $s$- or \didier{a} $p$-wave are emitted from point source A and the $s$-wave from point source B, respectively. \didier{By contrast, i}n figure \ref{fig:with_propagation}, either an $s$- or a $p$-wave is emitted from point source A, then propagates to point B and is finally scattered, the scattered $s$-wave \didier{being} emitted from point source B. 
As can be seen from Figures \ref{fig:cos-cos} and \ref{fig:cos-cos-1}, the difference between \didier{the} $s$-$s$ interference and \didier{the} $p$-$s$ interference appears as a cosine envelope, which makes the number of peaks/valleys differ by one, due to the \didier{change of } sign of the $p$-wave 
at $\theta=\pi/2$.
This is the reason 
\didier{why} the numerator of the first term in equation \ref{eq:Young} is $\Delta N$, while 
\didier{it is } 
$(\Delta N -1)$ \didier{in equation \ref{eq:new_eq}}. 
In the high energy regime, $\Delta N$ becomes so large that $\Delta N -1\simeq \Delta N$ and the energy-dependent parameter $\beta$ becomes negligible, 
\didier{making} our new formula for $p$-$s$ interference converge to the ordinary Young’s formula for $s$-$s$ interference.

\begin{figure}[htb]
\centering
\includegraphics[width=0.6\linewidth]{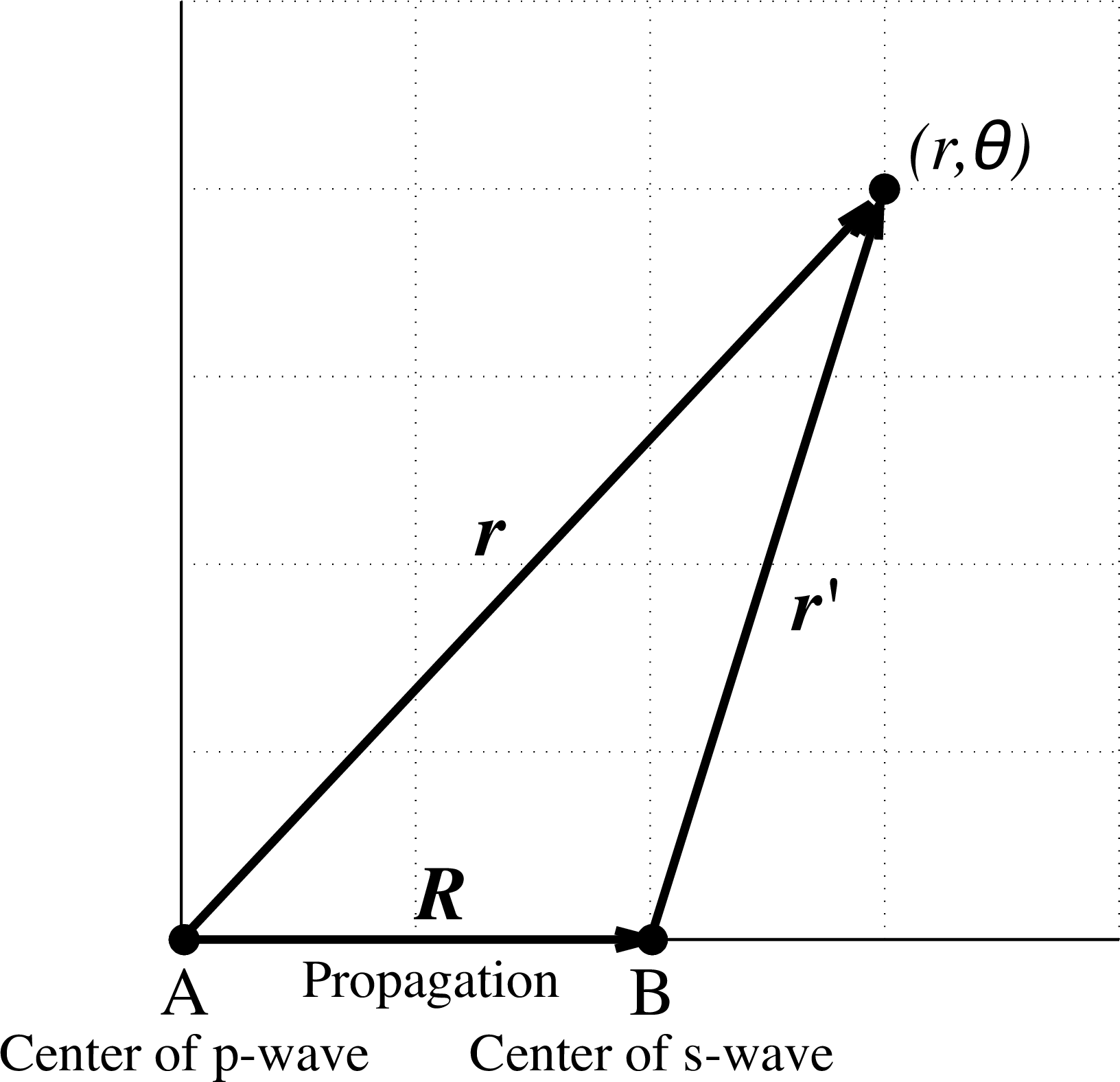}
\caption{\label{fig:model_p-s}
Schematic diagram of \didier{the} $p$-wave and $s$-wave interference. A $p$-wave is centered at  point A, an $s$-wave is centered at  point B, and \didier{the} vector \didier{connecting}  point A to  point B is defined as ${\bf R}$.
\didier{The} vector\didier{s} from point A and \didier{point} B to the observation point $(r,\theta)$ are defined as ${\bf r}$ and ${\bf r'}$ respectively.
}
\end{figure}

\ota{
\didier{Now,} we formulate the intensity observed at the point $(r,\theta)$ for the $p$-$s$ interference \didier {by} considering the propagation 
\didier{in the form of } a plane wave $e^{i{\bf k'}\cdot{\bf R}}$ 
(see \didier{the} schematic diagram of the model shown in figure \ref{fig:model_p-s}).
The intensity $I_{wave}(k,\theta)$ is given as 
\begin{eqnarray*}
I_{wave}(k,\theta)
&=\left|
 \frac{e^{ikr}}{r}\cos\theta
+e^{i{\bf k'}\cdot{\bf R}}\frac{e^{ikr'}}{r'}
\right|^2 \\
&=
\frac{\cos^2\theta}{r^2}
+\frac{1}{{r'}^2}
+2\Re\left(
\frac{e^{i(kR+kr'-kr)}}{rr'}
\cos\theta
\right).
\end{eqnarray*}
In \didier{the} limit $r\gg R$ and $r'\gg R$,
\begin{eqnarray*}
{r'}
=\sqrt{r^2+R^2-2rR\cos\theta}
\sim 
r-R\cos\theta,
\end{eqnarray*}
\didier{and} the intensity of \didier{the} interference of \didier{a} $p$-wave and  \didier{an} $s$-wave \didier{separated} by $R$ reduces to
\begin{eqnarray}
I_{wave}(k,\theta) &\sim
\frac{1}{{r}^2}
\left(
\cos^2\theta+1
+2\cos(kR(1-\cos\theta))
\cos\theta
\right).\nonumber\\
\end{eqnarray}
Thus, the fringes caused by \didier{the} $p$-$s$ interference are given by the last term $\cos(kR(1-\cos\theta))\cos\theta$
which agrees with the expression for the PA-MFPADs interference term given in equation~\ref{eq:F_theta}, except for the phase of the scattering amplitude. Therefore, the fringes of PA-MFPADs can be interpreted as the result of interference\didier{s} between the $p$-wave, which is excited along the molecular axis and propagates as a plane wave to the neighboring atoms, and the scattered $s$-wave, whose phase is shifted by $\phi(\theta)$.
}
{
}
%
%
%
%
%
\section{Conclusions}
We \didier{have} derived a new 
\ueda{formula, equation \ref{eq:new_eq}, 
\didier{to describe} the interference pattern\didier{s} that appear as \didier{a} {\it flower shape} \didier{in} PA-MFPADs of hetero-diatomic molecules} within the framework of Multiple Scattering theory, using \didier{(i)} the single scattering approximation, \didier{(ii)} the Plane Wave approximation, and \didier{(iii)} the Muffin-tin approximation.
\par
As the new equation was derived by assuming $\left|\left(kR-\frac{d\phi(\theta)}{d\cos\theta}\right) \cos \theta \right| \gg 1$, it works well for fringes close to the forward or \didier{the} backward direction. Thus, choosing two angles, one at the fringe nearest to the forward direction and the 
\didier{second at the} last fringe \didier{closest} to the backward direction, \didier{we are able to} reduce \didier{the} theoretical errors in our approximation \didier{of} the phase of \didier{the} scattering amplitude. 
%
\ota{
For practical use, the choice of peaks/valleys \didier{in} our new formula is one of the key points for \didier{a secure} analysis of experimental results.
  As can be seen from figure~\ref{fig:peaks}, the peaks and valleys appearing near the molecular-axis perpendicular direction 
  \didier{have a more complex structure than }
those appearing in the forward or backward direction.
Furthermore, as the photoelectron kinetic energy or bond length increases, the number of peaks and valleys increases \hatada{(see equation~\ref{eq:F_theta})}, which makes it difficult to distinguish fringes from each other \didier{for} middle\didier{-range} angle\didier{s}. For this reason, \didier{the} choice of two angles
,
one at the fringe nearest to the forward direction and the 
\didier{second at the} last fringe \didier{closest} to the backward 
direction,
\didier{allows} us to perform \didier{the} analysis with less experimental ambiguity.}
\par
The accuracy of this new bond length prediction equation was \didier{benchmarked} against PA-MFPADs of CO$^{2+}$ molecule calculated theoretically 
\didier{using} the Full-potential method. 
We found the relative error \didier{to be} surprisingly very small, i.e., less than 5\%. This encourages us to utilize this new formula to obtain bond length information of hetero-diatomic molecules on dissociation dynamics through time-resolved PA-MFPADs using the COLTRIMS-Reaction Microscope and \didier{the} two-color XFEL pump-probe set-up.
\par
Thanks to Multiple Scattering theory, the \didier{physical mechanism behind the} {\it flower shape} pattern \didier{in} PA-MFPADs 
\didier{has been} revealed, and the errors \didier{originating from the use of} \didier{the} Young's formula \didier{(}equation~\ref{eq:Young's_eq}\didier{)} to \didier{model} 
the {\it flower shape} \didier{has been} clarified. We 
\didier{have} pointed out that \didier{the} additional node 
\didier{appearing} in the {\it flower shape} \didier{is} due to the reversal of the sign of the direct dipole wave, i.e. \didier{a} $p$-wave from absorbing atom at $\theta=90^{\circ}$. Because of this effect, we should have an error in the parameter $\mu$ in Young's formula in equation~\ref{eq:Young's_eq} when $\theta_{\nu}<90^{\circ}$ and $\theta_{\nu+\mu}>90^{\circ}$, and we should subtract \hatada{$\pi$ in the numerator. }
In addition, the parameter $\beta$, which is not included in Young's formula, needs to be introduced.
This parameter $\beta$ depends on the kinetic energy of the photoelectrons. \didier{It} is not \hatada{sensitive to} 
\didier{the} bond length, and 
\didier{contains} information on the difference in phase of the scattered waves at the angles of the two selected fringes.
\par
In nature, a similar phenomenon of interference between $p$-waves and $s$-waves occurs within wave mechanics. We \didier{believe} that the newly derived \ueda{formula} for $p$-wave and $s$-wave interference\didier{s} \didier{can be} applied not only for PA-MFPADs but also in other fields of \didier{science}.
\section*{Acknowledgement\didier{s}}
This work was performed under the Cooperative Research Program of ``Network Joint Research Center for Materials and Devices''. K.H. acknowledges funding by JST CREST Grant No. JPMJCR1861 and JSPS KAKENHI under Grant No. 18K05027 and 17K04980. K. Y. is grateful for the financial support from Building of Consortia for the Development of Human Resources in Science and Technology, MEXT, and JSPS KAKENHI Grant Number 19H05628.
\section*{References}
\bibliographystyle{unsrt}
\bibliography{fukiko}
\end{document}